\documentstyle[twoside,fleqn,myespcrc2,epsf]{article}

\newcommand{\AmS}{{\protect\the\textfont2
  A\kern-.1667em\lower.5ex\hbox{M}\kern-.125emS}}

\newcommand{\vev}[1]{\langle #1 \rangle}

\def\lsim{\raise0.3ex\hbox{$<$\kern-0.75em\raise-1.1ex\hbox{$\sim$}}}
\def\gsim{\raise0.3ex\hbox{$>$\kern-0.75em\raise-1.1ex\hbox{$\sim$}}}

\title{
\vspace{-4.5cm}
\begin{flushright}
{\normalsize
\vspace{-.2cm}
September 1999}
\end{flushright}
\vspace{1.5cm}
Lattice QCD with Domain-Wall Fermions
\thanks{Talk presented by T.~Izubuchi at Lattice 99, Pisa, Italy.}}

\author{
Sinya Aoki $^{\rm a}$, Taku Izubuchi
\address{Institute of Physics, University of Tsukuba, Ibaraki 305-8571,
Japan},
Yoshinobu Kuramashi
\address{Department of Physics, Washington University, 
St. Louis, Missouri 63130, USA}
and
Yusuke Taniguchi $^{\rm a}$
}
\begin{document}

\begin{abstract}
We study the quenched lattice QCD using domain-wall fermions 
at $\beta=6.0$. 
Behaviors of both pion mass and the explicit breaking term
in the axial Ward-Takahashi identity support
the existence of the chiral zero modes.
We observe a good agreement between the pion decay constants
$f_\pi$ from both the conserved axial current and
the local current perturbatively renormalized at 1-loop.
Finally the possible existence of the parity broken phase
is also examined in this model.
\end{abstract}

\maketitle
\section{Introduction}
Domain-wall QCD (DWQCD)\cite{FS,Blum} is considered to have
good properties such as exact chiral symmetry without doublers, 
no $O(a)$ scaling violation and the existence of conserved axial current.
There exist several pilot studies\cite{Blum}, which seems to support
these superior properties.

We study DWQCD in the quenched approximation.
First we investigate the pion mass and the explicit breaking term of
the axial Ward-Takahashi identity to confirm the existence of 
the chiral zero mode at zero current quark mass.
Next we calculate the pion decay constant $f_\pi$ 
from both the conserved axial current and the local axial current,
using the perturbative renormalization factor\cite{BinRenorm} for
the latter, in order to check the reliability of the lattice
perturbation theory.
Finally we explore negative $m_f$ region to examine the existence of
the parity broken phase predicted in \cite{PBP}.

\section{Chiral symmetry}
The fermion action is identical to the original one\cite{FS}, 
with domain wall height $M$, 
bare quark mass $m_f$ and the extent of the 5th dimension $N_s$.
We employ 10--30 gauge configurations,
generated by the plaquette action at $\beta=6.0$ 
($ a^{-1} \sim 2$ GeV) on $16^3\times 32\times N_s$ lattices. 
The unit wall source without gauge fixing is used for quark propagators.
The mean-field estimate for the optimal value of $M$,
$M=1+4*(1-u)$ with $u=1/(8 K_c)$, gives $M=1.819$,
from $K_c=0.1572$ for the Wilson fermion at $\beta =6.0$.

\begin{figure}[t]
  \epsfxsize=7cm \epsfbox{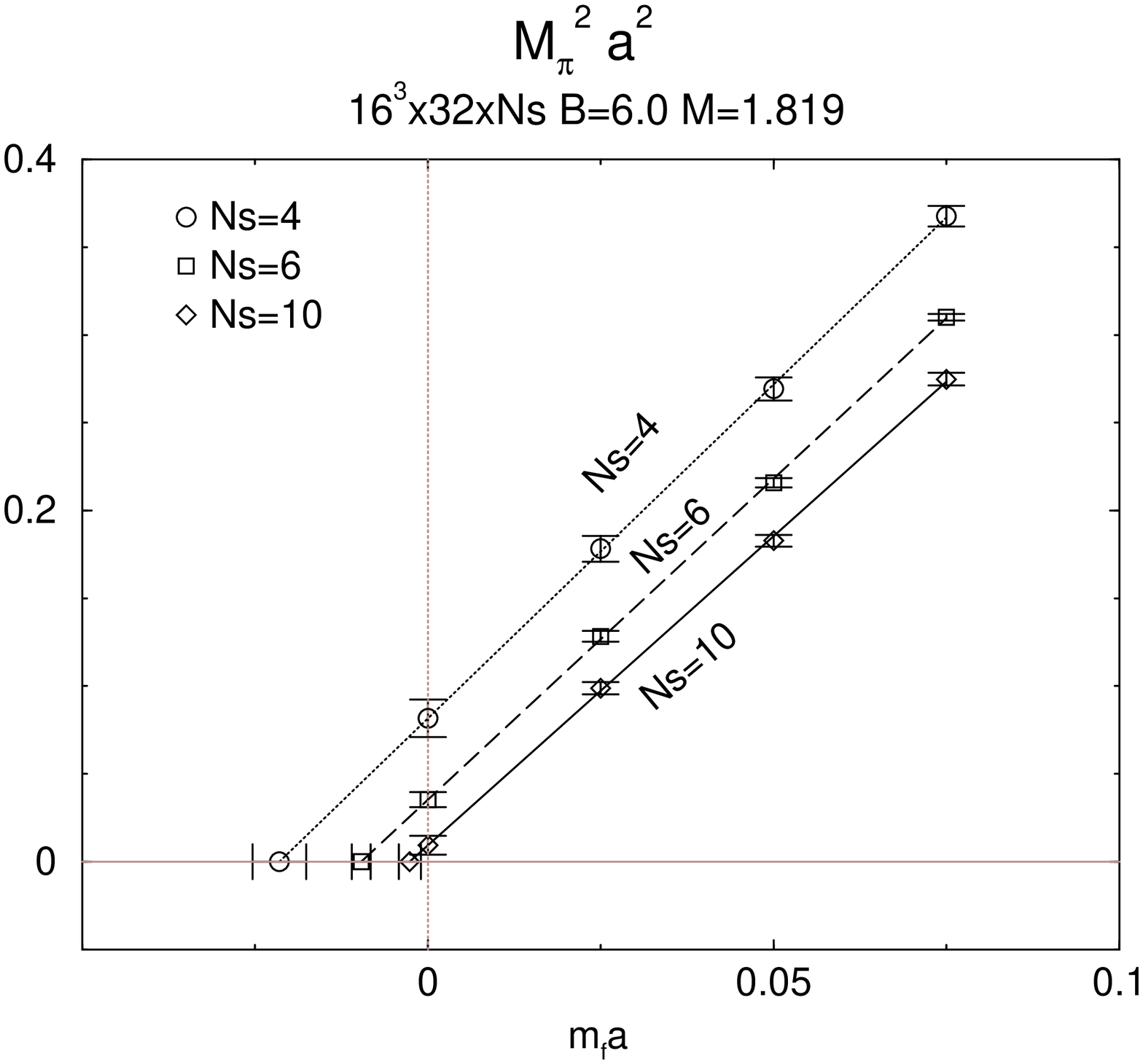}
  \vspace{-28pt}
\caption{Pion mass squared as a function of $m_f a$ 
at $M=1.819$ and $ N_s=4, 6, 10$. Solid lines show linear fits.
}
  \label{fig:Mpi2_mf}
\vspace{-22pt}
\end{figure}

First we investigate the existence of the chiral zero mode 
in the chiral limit of the model, $m_f\to 0$ and $N_s\to\infty$.
The pion mass squared is plotted as a function 
of $m_f a$ in Fig.\ref{fig:Mpi2_mf}. 
Since the linearity of $M_\pi^2$ in $m_f$ is well satisfied,
we linearly extrapolate it to $m_f=0$ for each $N_s$.
We also evaluate a critical quark mass $m_c(N_s,M)$ at which
the pion mass squared vanishes.

\begin{figure}[t]
  \vspace{-18pt}
  \epsfxsize=7cm \epsfbox{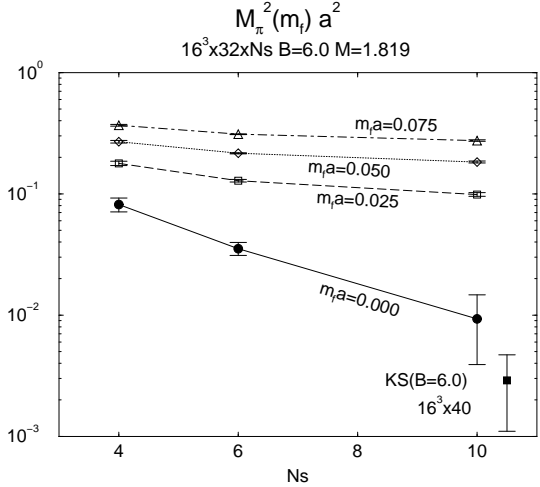}
\vspace{-28pt}
\caption{Pion mass squared as a function of $N_s$ 
at $M=1.819$. 
Filled square shows the value for the NG pion of
the KS fermion at $\beta=6.0$ on a $16^3\times40$ lattice.
}
  \label{fig:Mpi2_Ns}
\vspace{-22pt}
\end{figure}

\begin{figure}[t]
\vspace{-15pt}
  \epsfxsize=7cm \epsfbox{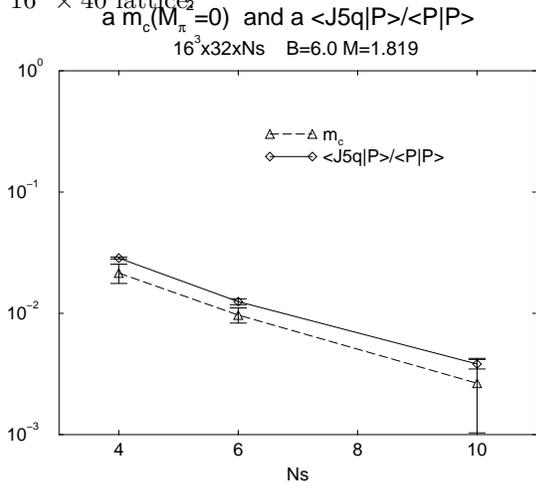}
 \vspace{-28pt}
\caption{critical quark mass $m_c$ and 
WI-mass $\vev{J_5^q|P}/\vev{P|P}$ 
extrapolated  to $m_f=0$ as a function of  $N_s$ at $M=1.819$ . }
  \label{fig:ItcX_Ns}
\vspace{-22pt}
\end{figure}

In Fig. \ref{fig:Mpi2_Ns}, $M_\pi^2$ is plotted as a
function of $N_s$ for $m_f a =$0.025, 0.050, 0.075 and $\rightarrow 0$.
Extrapolated values of $M_\pi^2$ at $m_f=0$ seem to vanish
exponentially in $N_s$, while $M_\pi^2$ at finite $m_f$  remains non-zero.
For $N_s=10$, $M_\pi^2(m_f=0)$ is already as small as that for the NG pion 
of the KS fermion at the same $\beta$ for the same spatial lattice size.
Furthermore $m_c(N_s, M=1.819)$  and the WI-mass\cite{FS}, defined by
$\vev{J_5^q|P}/\vev{P|P}$, also decrease exponentially in $N_s$,
as shown in Fig. \ref{fig:ItcX_Ns}. 
All these facts indicate that the chiral symmetry is restored for $m_f\to 0$
and $ N_s\to\infty$ at $\beta=6.0$.

\begin{figure}[t]
  \epsfxsize=7cm \epsfbox{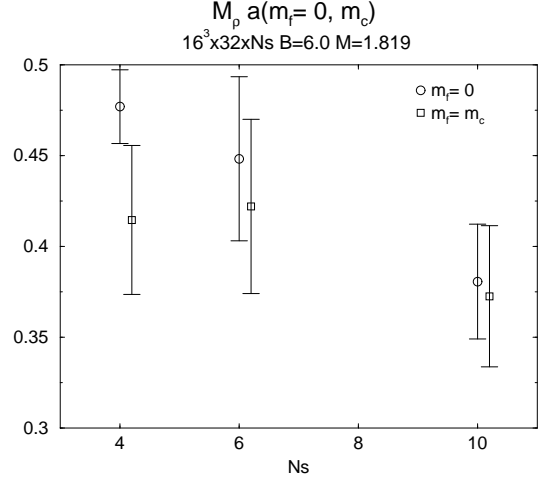}
 \vspace{-28pt}
\caption{Rho meson mass extrapolated  to $m_f=0$ and $m_f=m_c$
as a function of $N_s$ at $M=1.819$ . 
}
  \label{fig:Mrho_Ns}
 \vspace{-22pt}
\end{figure}

The lattice scale $a$ is set by the $\rho$ meson mass.
In Fig. \ref{fig:Mrho_Ns}, $M_\rho a$  is plotted as a function
of $N_s$. Circles  show linearly extrapolated values to
$m_f=0$, while squares show those to $m_f=m_c$.
For small $N_s(=4)$ the two ways of extrapolation give different
results, while two extrapolations are almost identical to each other
for $N_s=10$.
We also find that
the dependence of $M_\rho$ on the domain-wall height is mild for
$M=$1.7--1.9 and $N_s=10$: within statistical errors
the extrapolated values are consistent with each other.

\section{Pion decay constant $f_\pi$}
The pion decay constant is defined as 
$m_\pi f_\pi/Z_A = \vev{0|A_4|\pi}$, which is obtained 
from correlation functions of pseudo scalar density $P(t)$ 
and axial current $A_\mu(t)$ at zero momentum:
$\vev{X(t)Y(0)}=C_{XY} G(t)$ with $G(t)=\exp(-M_\pi t)/(2M_\pi V_s)$
for $X,Y = P, A_4$.
For the renormalization factor for the axial current $Z_A$,
we take unity for the conserved current and use the 
value from the mean-field improved perturbation theory\cite{BinRenorm} 
for the local current.

\begin{figure}[t]
  \epsfxsize=7cm \epsfbox{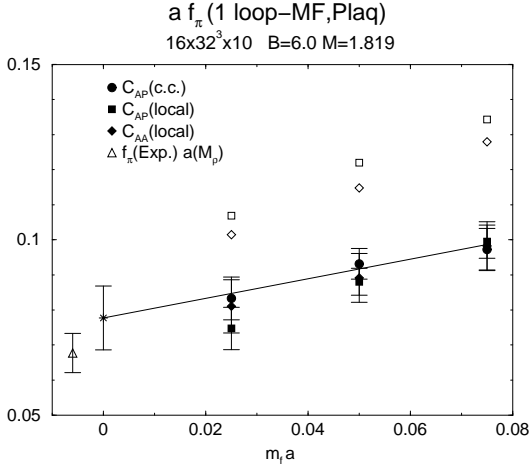}
\caption{$f_\pi$ in lattice unit as a function of $m_f a$
at $M=1.819$ and  $N_s=10$, together with  the experimental value. 
}
  \label{fig:FpiSRnmPlaq_M=1.819}
\end{figure}

In Fig. \ref{fig:FpiSRnmPlaq_M=1.819}, $f_\pi$ 
is shown as a function  of $m_f$.
Circles are obtained from $C_{AP}$ for the conserved
axial current, while squares from $C_{AP}$ and 
diamonds from $C_{AA}$ for the local axial current.
The triangle is the experimental value.
For local current, filled(open) symbols represent values with(without) 
perturbative corrections.
Three different estimates reasonably agree for all $m_f$ 
within current statistics, if 1-loop corrections are included.
It is also noted that
the value of $f_\pi$ from the conserved current, linearly extrapolated 
at $m_f=0$, is close to the experimental value.

\section{Parity broken phase}
\begin{figure}[t]
  \epsfxsize=7cm \epsfbox{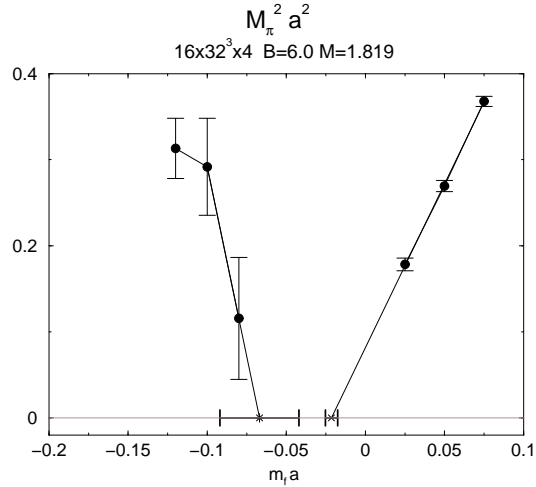}
\caption{$M_\pi^2$ in lattice unit as a function of $m_f a$
at $M=1.819$ and  $N_s=4$. 
}
  \label{fig:Mpi2_mfNeg}
\end{figure}

For finite $N_s$, the parity broken phase may exist 
in negative $m_f$ regions. 
As $N_s$ increases, the broken phase shrinks rapidly, and
the phase boundary, where the pion mass vanishes, 
converges to $m_f=0$\cite{PBP}.
To examine this parity broken picture in DWQCD,
we calculate pion masses at $m_f a = -0.120, -0.100, -0.080$ 
for $N_s=4$ and $M=1.819$.
The pion propagators at these parameters form peculiar shapes similar to
``W'' character, which has been often observed for the Wilson fermion 
near or in the parity broken phase.
Pion mass squared as a function of $m_f$ is shown
in Fig. \ref{fig:Mpi2_mfNeg}. 
Extrapolations of $M_\pi^2$ to zero both from positive and negative $m_f$ 
(two largest $m_f$ are used for negative $m_f$) indicate
that a parity broken phase may exist around $m_f a \sim -0.03$ at this
parameter. Needless to say, 
more high statistics and variation of parameters are necessary
for the definite conclusion.

\section{Conclusions and discussions}
At $\beta=6.0$ we have several indications that the chiral symmetry 
is restored at $N_s\to\infty$. 

We see that all three different estimates for $f_\pi$ are consistent with 
each other. This shows the mean-field improved perturbation theory works
well at $\beta =6.0$ in DWQCD. 
Although the value of $f_\pi$ at $\beta=6.0$ turns out to be
compatible to the experimental one, detailed scaling studies
at different $\beta$'s are needed before making the firm statement.
The chiral symmetry in DWQCD, however, may fail to
recover on the coarse lattice\cite{CPDWQCD}. If this is true,
the scaling studies of DWQCD become rather difficult.

We examine negative $m_f$ to see the parity broken phase in DWQCD. 
The result seems consistent with the parity broken picture,
though further confirmations for this are definitely required.

This work is supported in part by the Grants-in-Aid
of Ministry of Education
(Nos. 02375, 02373). 
TI and YT are JSPS Research Fellows.


\begin{thebibliography}{9}
\bibitem{FS} V.~Furman and Y.~Shamir, Nucl.\ Phys.\ B439 (1995) 54.
\bibitem{Blum} {\it For a review, see}
  T.~Blum, Nucl.\ Phys.\ B (Proc. Suppl.) 73 (1999) 167.
\bibitem{PBP} P. Vranas and I. Tziligakis, J. Kogut, hep-lat/9905018, 
T.~Izubuchi and K.~Nagai, hep-lat/9906017.
\bibitem{BinRenorm} S.~Aoki, T.~Izubuchi, Y.~Kuramashi and Y.~Taniguchi,
Phys.\ Rev.\ D59 (1998) 094505.
\bibitem{CPDWQCD} CP-PACS Collaboration, talk by Y.~Aoki,
in these  proceedings.
\end{thebibliography}
\end{document}